# Single-Cell Proteomic Technologies: Tools in the quest for principles

Nikolai Slavov[1,2]

1. Departments of Bioengineering, Biology, Chemistry and Chemical Biology, Single Cell Proteomics Center, and Barnett Institute, Northeastern University, Boston, MA 02115, USA;

2. Parallel Squared Technology Institute, Watertown, Massachusetts 02472, United States;

Nikolai Slavov (0000-0003-2035-1820)

**Abstract**

Over the last decade, proteomic analysis of single cells by mass spectrometry transitioned from an uncertain possibility to a set of robust and rapidly advancing technologies supporting the accurate quantification of thousands of proteins. We review the major drivers of this progress, from establishing feasibility to powerful and increasingly scalable methods. We focus on the trade-offs and synergies of different technological solutions within a coherent conceptual framework, which projects considerable room both for throughput scaling and for extending the analysis scope to functional protein measurements. We highlight the potential of these technologies to support the development of mechanistic biophysical models and help uncover new principles.

# Table of Contents

# Introduction

In the past decade, single-cell analysis has been dominated by nucleic acid-based profiling – especially single-cell transcriptomics (scRNA-seq) and genomics – which has supported characterization of cellular heterogeneity and the discovery of novel cell states(1, 2). However, these nucleic-acid-focused approaches provide an incomplete picture of cell biology, as RNA levels are indirect proxies for protein activity or phenotype(3–6). In fact, protein degradation is the dominant factor setting the abundance of many proteins(7, 8). Furthermore, cellular behavior depends on post-translational modifications (such as regulatory proteolysis, binding interactions, subcellular localization) that can be determined only by direct protein measurements. Consequently, there is a growing recognition that single-cell analyses must expand beyond genomes and transcriptomes to include direct measurements of other biomolecules, including proteins(9).

Indeed, direct protein analysis at the single-cell level is essential for understanding molecular mechanisms and cellular phenotypes since they depend on protein abundance, enzymatic activity, subcellular localization, and conformational dynamics. Yet achieving comprehensive single-cell proteome measurements remains technically challenging. Traditional single-cell protein assays (e.g. flow/mass cytometry or DNA-barcoded antibody panels) usually detect a few dozen proteins per cell and often suffer from limited antibody specificity(10, 11). Moreover, protein concentrations span a vast dynamic range (from ~1 to $10^7$ copies per cell); it poses changes for all methods, especially for single-molecule proteomics(12) and limits the detection of many low-abundance proteins. These limitations underscore the critical need for continued methodological developments towards overcoming the formidable challenges of accurate, comprehensive, and scalable quantification of single-cell proteomes. Many types of approaches may contribute solutions towards this goal(13), and here we focus on mass spectrometry (MS) proteomic approaches that have already demonstrated strong performance(14, 15) and hold much promise for further advancements(16). Accordingly, this review will focus on the advances in MS proteomics that recently have made remarkable gains. We will also project avenues that offer the potential for substantial technological and methodological gains in the coming years.

The direct quantification of proteins at sufficient accuracy and throughput may enable inference of direct molecular mechanisms, and potentially new new principles. This appealing possibility is the focus of the last section that contrasts formal statistical inference based on associations with mechanistic models grounded in biophysical principles and direct molecular interactions. We emphasize why measuring proteins (rather than only transcripts) is crucial for causal and mechanistic understanding of biological systems. We consider current limitations and outline future directions for the field, including promising approaches to harness single-cell protein data for causal discovery.

# The arc of technology development

MS has been used to detect a few abundant proteins and peptides in individual cells for decades (reviewed in refs.(17, 18)), but until recently the sensitivity of MS was considered insufficient for the quantification of many proteins in single cells from complex tissues(19). Yet, sensitive sample preparation methods(20–22) and new experimental designs for multiplexed data acquisition(23, 24) started changing these perceptions. In the early stages, the use of isobaric mass tags together with carrier samples, introduced by the SCoPE-MS method(23), was very helpful to increase the copy number of fragmented peptides and thus support sequence identification given the more limited sensitivity of older instruments, such as Q-Exactive orbitraps. Isobaric tagging allows peptides from multiple cells to be measured in one MS run, while the pooling of peptide fragment ions across many isobarically labeled single cells and a "carrier" (a larger sample) increases the ions supporting sequence identification(25–28). The convincing demonstrations of the feasibility of single-cell proteomics analysis by MS (reviewed in ref.(10, 29–31)) accelerated the development of both MS instrumentation and methods that are reviewed in the sections below.

The development of these technologies was fueled by considerable technological and conceptual opportunities(32). A notable advantage of proteins is that they have about 10,000-fold higher copy numbers per cell than transcripts, which allowed even early generation methods to quantify proteins based on better counting statistics than parallel transcriptomic analysis(26). These technical advantages and the biological imperatives for direct quantification of proteins supported the development of single-cell proteomics. New MS instruments(33, 34) enabled deeper proteome coverage(15, 35), though older instruments continue to support biological investigations, often aided by isobaric carriers(27, 36–39) and intelligent data acquisition(40, 41). This progress happened in the context of the larger and more mature single-cell transcriptomic field, which provided both directions by analogy and high bars to meet and exceed for a nascent field with limited funding support. The parallel developments of other single-cell omics and spatial methods (whose spatial resolution relies on the sensitivity improvements of MS) has provided opportunities for cross pollination(42, 43).

# Sample preparation and peptide separation

One of the main challenges in the field has been to develop methods for sample preparation that simultaneously minimize protein losses and contaminants while preparing many cells in parallel. A variety of sample preparation methods have been developed, which can be grouped by the physical platform used, Figure 1. Here, we categorize these methods into four platforms – **multiwell plates**, **microchips**, **glass slides**, and **microfluidic systems** – and describe representative techniques, core features, and advantages of each. We specifically highlight how each platform addresses key challenges: minimizing protein loss, reducing contamination, and enabling parallel processing, in both label-free and multiplexed MS contexts. A few methods, such as cell lysis by sonication(28), acoustic levitation(44), sampling by patch-clamp(45) and microsampling(46), fall outside of these major categories and will not be discussed here.

**Categories of sample preparation methods for single-cell proteomics**

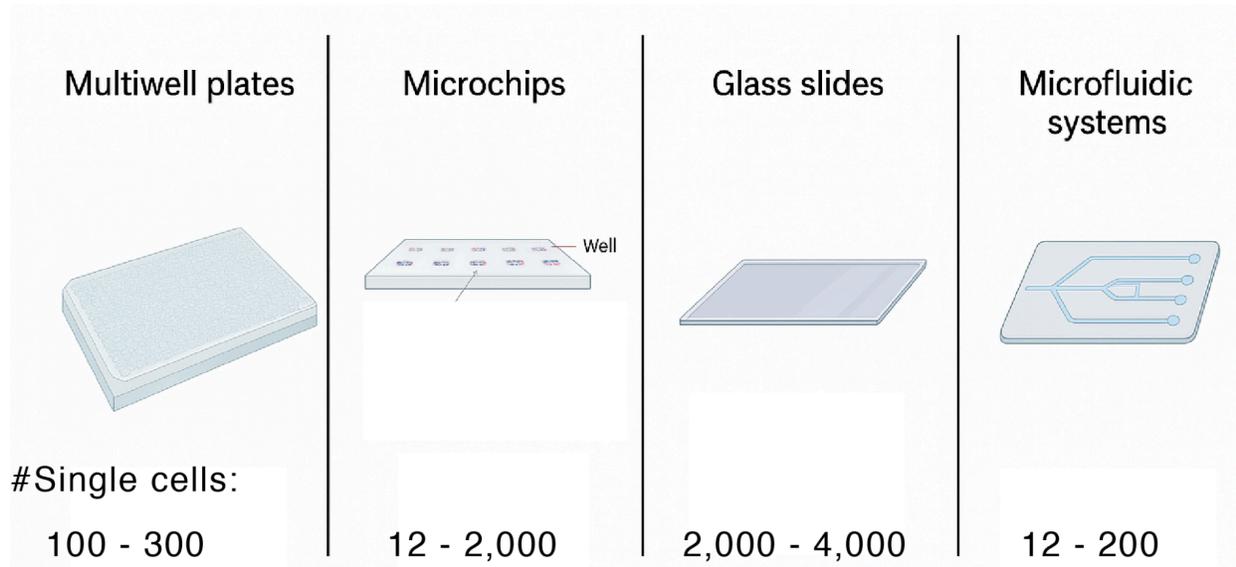

Figure 1. Sample preparation methods for single-cell proteomics categorized by the physical platforms used. The number of single cells that can be prepared in parallel as part of a single batch is listed at the bottom.

## Multiwell Plate-Based Methods

Multiwell plates (e.g. 96- or 384-well plates) provide an accessible platform for single-cell proteomics sample preparation. Typically, single cells are isolated (via FACS, CellenONE or micromanipulation) into individual wells, and all proteomic processing – cell lysis, protein digestion, and optionally isobaric labeling – is carried out in the same well. Volumes are kept low (on the order of 1–10 μL) to reduce surfaces and dilutions that could cause protein losses. The first such example is the Minimal ProteOmic sample Preparation (mPOP) protocol(22, 47), which employs a freeze–heat lysis in pure water within 384-well plates. This gentle one-pot lysis avoids detergents or chaotropes, thereby eliminating the need for cleanup and minimizing surfaces that peptides contact, which in turn reduces adsorptive loss and risk of contamination. After lysis and tryptic digestion, peptides from single cells can be chemically labeled with mass tags (e.g., TMTpro, mTRAQ, dimethyl, PSMtags) directly in the wells. Numerous other methods have built upon mPOP and use multiwell preparation methods(34, 48–50). These methods are commonly used for label-free single-cell proteomics though some can support labeling.

**Core features and advantages:** Multiwell plate methods leverage standard labware and liquid handling, making them relatively easy to implement. They are compatible with existing automation (e.g. robotic pipettors or acoustic dispensing) to handle small volumes and reduce human contact contamination. When optimized, all steps occur in a single well, which minimizes transfers between vessels that would otherwise cause protein loss. Multiwell formats inherently support parallel processing of a few hundred single cells. In summary, plate-based methods provide an accessible entry to single-cell proteomics with simple one-pot protocols that use µL-scale volumes, larger than other methods. Pooling of labeled cells generally remains a manual step (for multiplexing).

## Microchip-Based Methods

Microfabricated chip platforms improve on conventional plates by allowing to reduce the surface areas and volumes. These platforms typically consist of an array of small wells or chambers on materials like silicon or polymer, often with specialized coatings. Each well functions as a small reactor for a single cell's proteome, enabling one-pot processing. A landmark in this category is the Nanodroplet Processing in One Pot for Trace Samples (nanoPOTS) (51). NanoPOTS uses photolithographically patterned glass chips containing nanowells of ~200 nL or smaller, into which single cells are deposited and lysed. By confining the entire sample prep to small volumes, nanoPOTS demonstrated improved protein recovery and identification from as few as 1–100 cells(51). Its more recent iteration, the nested nanoPOTS chip (N2), further shrinks reaction volumes to <30 nL per well and packs more nanowells per chip(52). The N2 design "nests" multiple tiny wells under a larger droplet: after each single-cell digest is labeled (with TMT) in its nanowell, a microliter-scale droplet added on top simultaneously pools all wells in that cluster for easy collection. This innovation facilitates the pooling and reduces surface contact. Another representative is the proteoCHIP, a commercial micromachined polytetrafluoroethylene (PTFE) chip with arrays of conical wells (e.g. 96-well format)(53). The proteoCHIP is pre-filled with an inert oil overlay. After digestion (and labeling if multiplexing), the entire well content can be transferred for analysis. Both N2 and the proteoCHIP are operated with an automated nanoliter dispensing robot (CellenONE).

**Core features and advantages:** Microchip-based methods reduce volume compared to multiwell plate methods, which can help reduce protein losses and allow for higher concentrations that in turn improve reaction efficiency. Both label-free and multiplexed workflows are compatible: a proteoCHIP can be used in "LF" mode to prepare ~96 single-cell digests separately for individual nanoLC-MS runs, or in multiplexed mode to label and then pool cells (e.g. 16 single cells per set) within the chip. Likewise, nanoPOTS chips were first used for label-free profiling of small bulk samples composed of 10–100 cells, but later extended to TMT-labeling of single cells (N2 chip) supporting the quantification of about 1500 proteins per cell. The trade-offs include the need for specialized microfabricated consumables and sometimes custom robotic handlers.

## Glass Slide-Based Droplet Arrays

Glass slide-based methods increase the throughput and flexibility of sample preparation by using open surfaces to arrange nanoliter droplets, rather than fixed wells. In this approach (nPOP), microscope slides with special coating are used as a platform on which thousands of tiny reaction droplets are dispensed in an array. nPOP has allowed the parallel preparation of thousands of single cells by spotting nanoliter droplets on unpatterned slides(15, 54). In nPOP, an automated dispenser (CellenONE) deposits single cells in ~10–20 nL DMSO droplets onto the slide in prescribed patterns. Unlike microfabricated chips with fixed well locations, this open format offers spatial freedom to design droplet arrays of any size or grouping to suit the multiplexing scheme. For example, droplets can be arranged in clusters corresponding to an isobaric labeling set. Leduc *et al.* demonstrated a 29-plex design where 29 single-cell droplets (each destined for a different TMTpro tag) are placed in proximity as one cluster, enabling simultaneous preparation of 3,584 single cells across four slides in one nPOP run. Each cell-containing droplet undergoes lysis, proteolysis, and labeling on the slide surface. Because the slide is flat and unconfined, reagents can be added by the robot to each droplet in parallel (e.g. overlaying lysis buffer, then adding TMT reagents). After labeling, droplets belonging to the same multiplexed set are aspirated and combined (via the CellenONE system) for MS analysis, or in a label-free scenario, each droplet can be picked up individually for injection.

**Core features and advantages:** The glass slide platform offers maximal throughput and scalability with minimal reaction volumes. Slides are easy to obtain and prepare (no microfabrication needed), and hundreds to thousands of nanodroplet reactors can be arrayed on a single slide. The capacity is limited mainly by the dispensing speed and slide area – indeed, nPOP's throughput (over 3,700 single cells per prep) is currently constrained by the dispensing speed (~2 hours). This approach achieves the smallest reaction volumes (droplets in the tens of nL), which translates to lower surface contact and high protein recovery comparable to nanowell chips. An advantage unique to slides is the flexibility to adapt any plex size or experimental design without having to redesign a device. Researchers can simply program different droplet patterns (e.g. to use a 35-plex TMTpro kit or a 9-plex PSMtag set) on the same slide format. As with microchips, all reactions for a given cell occur in one droplet, obviating transfers and thereby limiting losses and contamination. The open format, however, requires careful environmental control – typically the slide is kept in a humidity-controlled enclosure to prevent evaporation and airborne contaminants from entering the tiny droplets. When properly controlled, slide-based nPOP has demonstrated competitive depth and quantitative accuracy: about 3,000–3,700 proteins quantified per human cell using a plexDIA (multiplexed DIA) workflow, with low measurement noise (e.g. blank droplets give very low signal) and high quantitative accuracy(15). While nPOP was introduced for multiplexed TMT/plexDIA processing, the protocol can be used for label-free preparation as well. In that case, many single cells are prepared in parallel. Then, they are individually collected for separate LC-MS runs, though the advantages of nPOP are most pronounced for multiplexed workflows. In summary, glass slide droplet arrays combine miniaturization with massive parallelisation, making them ideal for suitable multiplexed single-cell proteomics, provided one has access to the requisite dispensing and collection instrumentation.

## Microfluidic Integrated Systems

Microfluidic systems take a different approach by integrating the entire sample prep workflow within closed microdevices (Lab-on-Chip devices). These platforms use micro-scale fluid channels, chambers, and valves (often in polydimethylsiloxane, PDMS) or electrowetting-based digital chips to automate protocols on chip. The goal is to minimize exposure, thus reducing contamination, while precisely manipulating microliter or sub-microliter volumes to improve recovery. One implementation, SciProChip, uses a two-layer PDMS microfluidic chip, which features microvalve-controlled flow channels for cell isolation, lysis, digestion, on-chip solid-phase extraction cleanup, and collection – all on one device(55). Single cells are captured in dedicated microchambers, imaged/counted in situ, then lysed and digested in hundreds of nL reaction vessels; the resulting peptides are passed through an integrated C18 resin microcolumn for desalting before elution off-chip to MS. The entire process is coordinated by dozens of microvalves that route fluids from inlets to various chambers in a programmable sequence. Another approach is digital microfluidics, which uses electrostatic forces to manipulate discrete droplets on an array of electrodes(56). In this system, droplets containing single cells and reagents are moved, merged, and split on a planar chip under software control. All steps – cell lysis, reduction/alkylation, and trypsin digestion – are executed in succession by shuttling the droplet to different zones of the chip, with on-chip vision to ensure one cell per droplet. Microfluidic chips can also be used to sensitively enrich for post-translational modifications, such as phosphorylation(57).

**Core features and advantages:** Microfluidic platforms bring the benefit of full automation in a closed environment. By integrating all steps on a chip, they reduce the risk of external contamination (no open tube transfers or manual pipetting of tiny volumes). They also facilitate sample traceability since each cell stays in a known micro-chamber or droplet throughout processing. While microfluidic chips allow for processing multiple cells concurrently (SciProChip contains 20 independent single-cell processing units on one device), the parallelization demonstrated so far is substantially lower than with glass slides. Current demonstrations have mostly focused on label-free analysis – each cell's digest is delivered separately to MS, though in principle chip designs may be amenable to multiplexing. A challenge for microfluidics has been reagent compatibility since organic solvents for labeling can swell PDMS or interfere with droplet actuation. In sum, microfluidic systems offer a highly controlled, automated environment for single-cell proteomics though remain limited in accessibility,

## Minimizing artifacts

An important objective for all sample preparation methods is to minimize artifacts during sample preparation. These can include stresses that result in physiological changes during cell isolation or storage, cell permeabilization, variations in cell lysis or protein digestion efficiency. Following the community guidelines(58) and monitoring for batch effects can mitigate their impact on data interpretation. As the field matures, these factors have received more attention(42), with articles characterizing the impact of cell storage conditions(59) and suggesting fixation protocols for increasing the proteome stability during sample processing(60). A particular concern is the possibility of proteins leaking from the cells if cell integrity is lost during cell isolation from tissues. Since such protein loss affects some groups of proteins more than others, protein leakage systematically affects groups of functionally related proteins and may be mistaken for biological signals(61). Such artifacts may be avoided by using cell permeability dyes to isolate only intact cells during sample preparation or mitigated by using computational methods to remove strongly affected cells post data acquisition(61).

## Peptide separation

Handling the enormous dynamic range of proteomes requires high-performance separation of peptides(12). It allows for efficient sampling and quantification of analytes of vastly different abundance and achieving deep proteome and sequence coverage in a single experiment. This separation can be achieved in liquid phase (using chromatography or capillary electrophoresis) or in gas phase (using ion mobility). This section will focus on liquid phase separation (which is the most common and powerful form of separation) while gas phase separations will be covered in the next section on data accusation.

The sensitivity of single-cell proteomics can be increased by low-flow peptide separation techniques. Reducing flow rates into the low-nanoliter per minute range (via nanoLC or capillary electrophoresis separations) improves electrospray ionization efficiency, delivering more peptide ions. This is because tiny flow rates produce finer ESI droplets with less ion suppression. For example, dropping from 300 nL/min to 20 nL/min can increase signal intensity per molecule by roughly 5-10 fold(62). In practice, such ultra-low-flow setups translate to deeper proteome coverage from small samples(63) and single cells(64). Capillary electrophoresis (CE)–MS approaches, which inherently operate at sub-50 nL/min flows, have likewise shown sensitivity gains(65), especially for the nascent analysis of intact proteins(66). Such results demonstrate that miniaturized, low-flow separations can substantially improve ion utilization and signal intensity, directly translating into deeper single-cell proteome coverage.

These benefits, however, come with trade-offs in robustness, accessibility, and reproducibility(67). Operating narrow-bore LC columns (20–50 µm i.d.) and nanoflow regimes are technically challenging and prone to clogging. Gains from using narrower columns can be tempered by practical issues like sample loss and long loading times: nanoLC systems often require relatively large volumes and extra dead volume flushing, which at ~10–20 nL/min flow leads to dilute samples and delays, potentially undermining sensitivity gains. Capillary electrophoresis, while very sensitive, has its own reproducibility hurdles: peptides may adsorb to the capillary walls and slight drifts in electroosmotic flow or electrospray stability can erode run-to-run consistency. Moreover, CE-MS demands handling of picoliter–nanoliter sample volumes, which is non-trivial given that most single-cell sample prep methods output in the microliter range – introducing a tiny aliquot into the capillary without loss or contamination is a significant challenge. Consequently, both narrow-bore nanoLC and CE-based workflows have historically shown reduced robustness and accessibility, often confined to specialized labs with expert optimizations(67).

To integrate the strengths of low-flow methods while mitigating their limitations, ongoing efforts are focusing on new technologies and hybrid strategies. Researchers are developing alternative column formats – such as monolithic and nano open-tubular columns – that can sustain efficient separations at ultralow flows without the clogging and packing difficulties of conventional packed beds. Improved CE-MS interfaces are also being explored: for example, chemically coated capillaries (e.g. with linear polyacrylamide) to prevent analyte adsorption and stabilize flow, and on-line preconcentration techniques like sample stacking or solid-phase microextraction to increase injection capacity. In parallel, multiplexed low-flow LC systems are boosting throughput – notably, dual-column nanoLC setups increase the analysis throughput by decreasing wait times for sample loading and column regeneration(14, 68). Such innovations may capture the superior ionization efficiency of ultralow-flow LC and CE methods while overcoming their drawbacks.

The time needed to separate peptides determines the throughput of the analysis. Thus, multiple methods have focussed on reducing the duration of the separation and more recently on staggering the separation of multiple concurrent samples via a method termed timePlex(69). These approaches will be discussed in the section on increasing throughput.

# Data acquisition and analysis

The methods used for acquiring mass spectra for single-cell proteomics can be broadly divided into data dependent acquisition (DDA) and data independent acquisition (DIA). DDA methods were more common with the earlier approaches while DIA methods have increased in popularity over the last 5 years. DDA and DIA methods are also used for acquiring spectra from bulk samples and have been extensively reviewed in the literature, e.g., ref.(70). Here we will highlight only aspects directly pertinent to maximizing sensitivity for single-cell proteomics.

## Increasing the efficiency of ion utilization

Improving ion utilization efficiency can directly translate to enhanced sensitivity in single-cell proteomics. By maximizing the fraction of ions that are captured and fragmented, more peptide ions can be detected from each cell, increasing proteome coverage.

## Isolating ions with wide windows

One strategy is to use wider precursor isolation windows in DIA acquisitions, as implemented by Derks *et al.*, 2022 with the development of the plexDIA method(71). Wide isolation windows allow multiple precursors to be co-isolated and fragmented in parallel during DIA, thus increasing the total ions sampled per MS/MS scan(16, 72). This approach enables longer ion accumulation times and higher signal for low-abundance peptides, outweighing the increase in spectral complexity when sample complexity is lower. In practice, Derks *et al.* achieved single-cell analyses on Q-exactive classic using only four very wide isolation windows for MS2 scans, which permitted ~300 ms accumulation times, thereby supporting sensitive analysis on an MS instrument from 2010. Wider isolation windows have also been used with DDA to co-isolate multiple precursors simultaneously, thus increasing ion utilization and the number of identified peptides(73). A notable tradeoff of wide isolation windows is the reduction of the specificity of associating precursors and fragments, which adversely affects sequence identification. The optimal window width and other acquisition parameters may be optimized by open source data driven pipelines, such as DO-MS(74).

## Parallel ion accumulation and separation

A second, complementary strategy is the integration of ion mobility separation to improve ion utilization. Trapped ion mobility spectrometry (TIMS) coupled with parallel accumulation–serial fragmentation (PASEF) on timsTOF instruments increases duty cycle efficiency by concentrating ions and synchronizing their release for MS/MS(75). For example, slice-PASEF continuously scans the quadrupole across the mobility dimension to fragment most ions and increase ion usage in each TIMS frame(76). This method boosts sensitivity and proteome depth for tiny samples while allowing for a short duty cycle compatible with short separation times and thus more single cells analyzed per unit time. Increasing the resolution of ion mobility separation can further increase the efficiency of this category of methods(77).

## Efficient ion handling and detection

A third key strategy for increasing ion utilization – and thus sensitivity – is to improve the mass spectrometer's ion transmission and detection efficiency through advanced hardware design. This approach has been used by many instruments (as discussed below) and is particularly emphasized with the Orbitrap Astral mass spectrometer that significantly reduced losses from ion injection to detection. Its dual-stage analyzer (Orbitrap plus the high-speed Astral analyzer) uses enhanced ion optics (e.g. high-capacity funnels and optimized quadrupoles) and a parallelized acquisition scheme to achieve Active Ion Management with highly efficient ion transport(33). This means a greater proportion of ions generated from a single cell reaches the detector. As a result, even narrow window DIA (which by design filters out most ions) can identify over 5,000 proteins from 250 pg of HeLa digest(35, 78). Wider isolation windows further improve the efficiency of ion utilization and increase proteome coverage.

## Intelligent data acquisition

Real-time control of MS acquisition has contributed to single-cell proteomics by intelligently guiding data acquisition to increase peptide identification, data completeness and depth. Traditional topN data-dependent acquisition stochastically selects the most intense precursors, often missing lower-abundance yet biologically relevant peptides and yielding run-to-run variation. Real-time acquisition strategies mitigate this limitation. For example, MaxQuant.Live interfaces with Orbitrap instruments to detect thousands of predefined precursors by adapting the acquisition parameters for MS/MS based on the measured retention times(79). As another example, Real-Time Search (RTS) on Thermo Tribrid Orbitraps uses an on-the-fly database search of MS2 spectra to guide acquisition(80). In an RTS-enabled method, the mass spectrometer's linear ion trap can rapidly identify a peptide sequence during acquisition; If a confident ID is found, the instrument triggers advanced scans (such as an Orbitrap MS2 or MS3) for quantification when using isobaric mass tag, such as TMTpro. These real-time control strategies markedly improve the efficiency of MS, both for bulk and single-cell proteomics.

Prioritized Single-Cell Proteomics (pSCoPE), developed by Huffman *et al.* (2023), introduced a multi-tier prioritization schema implemented by real-time retention time alignment and control of data acquisition(40). In pSCoPE, each peptide from an inclusion list is assigned a priority level, so that high-priority precursors are consistently selected for MS/MS in every single-cell run, and accumulated for longer if deserved for increased sensitivity. This real-time prioritization led to more consistent peptide identification across cells (greater data completeness) and doubled the proteins quantified per cell compared to standard methods. Notably, instrument time is allocated to peptides that are identifiable (or biologically relevant), improving proteome coverage and sensitivity. The prioritization can even allocate longer ion injection times to low-abundance priority peptides, increasing their MS2 signal and further increasing sensitivity and depth. pSCoPE is relatively accessible since it's implemented by free software and can be used with many Thermo instruments, including all Q-Exactive and Exploris instruments. When used on a Q-Exactive classic instrument, pSCoPE allowed quantifying about 1,500 proteins per human cell while achieving over 90% data completeness across many single cells(40). Beyond the initial report, pSCoPE has been used to analyze single cells from primary mouse and human tissues(5, 61).

Other approaches for intelligent data accusation use real-time search on Thermo Tribrid instruments. By performing an immediate database search on each MS2 spectrum, the RTS-enabled Orbitrap Eclipse only triggers an MS3 scan (with synchronous precursor selection, SPS) if the peptide was identified in real time. This approach curtails co-fragmentation interference and ratio compression: only fragment ions from the identified peptide are used in the MS3 quantification, minimizing contamination from co-isolated species. Furtwängler *et al.* demonstrated that an RTS-SPS-MS3 method on a Tribrid yields higher quantitative accuracy than classical MS2, while a related RTS-based MS2 strategy (termed RETICLE) achieved >1,000 proteins quantified per cell(81).

In summary, real-time control of data acquisition by prioritization and Thermo RTS both exemplify how intelligent, on-the-fly control of MS acquisition refines peptide selection, increases single-cell proteomic depth, and diminishes interference for more reliable quantification. These approaches are particularly advantageous when multiplexing single cells with isobaric mass tags, e.g., TMTpro. Other approaches for real-time intelligent data acquisition have been reviewed in ref.(82). While applicable to single-cell proteomics in principle, they have not yet been applied. Some approaches have been applied to DIA(83) and specifically optimized for single cells(84).

## Using carriers

The limited copy number of proteins present in single cells pose challenges to sequence identification. This challenge can be mitigated by using isobaric or isotopologous carriers. These are larger proteome samples, sometimes spiked peptides, that are labeled by mass tags and combined with single cells labeled by mass tags. Isobaric carriers use isobaric mass tags and enable the pooling of peptide fragments from the carrier and the single cells in each fragmentation spectrum, which provides more fragments for sequence identification(25, 28). As discussed in other sections, isobaric carriers helped establish the feasibility of single-cell proteomics by MS(85). Isotopologous carriers rely on the coelution and the known mass offset between the carrier and the single-cell peptides to propagate sequence identifications(71, 86). They have been used both with bulk samples(87) and with single cells and single nuclei(88, 89). The amount of both isobaric and isotopologous carriers needs to be balanced according to the experimental objectives(25) to mitigate adverse effects in quantification(90) and according to the community guidelines(58).

## Technological advances of instrumentation

Recent hardware innovations have substantially improved single-cell proteomics sensitivity by maximizing ion transmission, handling, scanning speeds, and ion detection. Refinements in nano-electrospray ion sources now produce significantly more peptide ions from picogram-scale samples. These brighter ion sources, when paired with enhanced ion optics (such as multi-stage ion funnels and high-transmission ion guides), reduce ion losses between the source and analyzer.

For example, the Bruker Captive Spray Ultra source (with a larger capillary and optimized vortex gas flow) and updated ion funnel interfaces significantly increase ion transfer efficiency. Concurrently, faster and more parallel data acquisition architectures have raised MS/MS scanning speed without sacrificing sensitivity. The timsTOF Ultra's trapped ion mobility spectrometry (TIMS) tandem-MS system exemplifies this: a dual-TIMS design continuously accumulates ions in one chamber while releasing them from a second, achieving highly efficient duty cycle via Parallel Accumulation–Serial Fragmentation (PASEF). Coupled to a fast time-of-flight analyzer and advanced 14-bit digitizers, this allows high MS/MS scan rates that facilitate high peptide sampling density across chromatographic peaks. Moreover, integrating ion mobility separation not only boosts duty cycle but also adds an extra dimension of gas-phase separation (by collisional cross-section), which reduces spectral complexity and improves detection of low-abundance peptides. These hardware advances have translated into deeper proteome coverage from small inputs. For example, recent label-free single-cell runs on the timsTOF Ultra 2 achieve on the order of 4,000 - 5,000 protein groups identified per Hela cell. These improvements make feasible analysis of single organelles and much smaller primary single cells, albeit with much lower proteome coverage, thereby delivering on the biological promises of the technology(5, 37, 89, 91).

As another example, the Orbitrap Astral mass spectrometer enables parallelized and fast scans by using a dual-analyzer design featuring an Orbitrap (for MS1) coupled to a new high-speed Astral time-of-flight analyzer(33). The Astral analyzer uses an asymmetric ion flight path and a dual-pressure linear ion trap (ion processor) to achieve high sensitivity and 200 Hz MS/MS throughput at low ion accumulation times. This parallelized ion handling pipeline allows multiple ion packets to be processed simultaneously (two in the ion trap, one in the multipole, and others being detected in Orbitrap and Astral), effectively aligning the Astral's 200-Hz MS/MS acquisition with adequate ion cooling time. Crucially, the efficient ion transfer and extended flight path of the Astral analyzer preserve ion signals across a wide dynamic range. By decoupling full scans and fragment scans between two analyzers, the Orbitrap Astral maximizes both sensitivity and speed – enabling high proteome depth from ng and pg level samples(33). Recent studies report that the Orbitrap Astral can identify about 5,000 protein groups from single Hela cells in label-free analyses(35, 78).

## Analysis of mass spectra

Identifying and quantifying peptides from mass spectra acquired from single cells has relied mostly on the same algorithms and software tools as those used with bulk samples(92–95). These tools have been extensively benchmarked and reviewed in the literature and will not be reviewed here. Rather, we will focus on methods more specific to single-cell proteomics.

One such group of methods increases data completeness by propagating peptide identifications from single cells (or reference/carrier samples) where a peptide is confidently identified to cells where it is not. For example, DART-ID was specifically motivated by single-cell proteomics data and implements Bayesian frameworks for global retention time alignment and for incorporating these estimates towards improved confidence estimates of peptide spectrum matches(96). Similarly, IceR(97) and IonQuant(98) can propagate peptide sequence identification and improve proteome coverage and data completeness in single-cell proteomics. Despite using such methods, low data completeness often remains a challenge in single-cell proteomics data and needs to be addressed by downstream computational methods(99). This downstream analysis remains underdeveloped in the context of single-cell proteomics and is beyond the scope of this review. We will just note that missing data leads to fundamental uncertainty in the results; this uncertainty needs to be propagated and reflected in the conclusions, for example by using multiple imputation approaches(58).

# Increasing throughput

Major approaches for increasing the throughput of single-cell proteomics analysis include (i) shortening of peptide separation times, (ii) multiplexing single cells, and (iii) multiplexing peptide isolation and fragmentation for MS2 analysis as performed with DIA and wide window DDA analysis. Based on the degree of multiplexing, methods can be classified as not multiplexed (label free DDA), multiplexing only samples (TMT-DDA), multiplexing only peptides (label free DIA) and multiplexing both samples and peptides (plexDIA and timePlex), as shown in Figure 2. The sections below briefly outline these methods within a coherent conceptual framework.

## Short separation times

One straightforward approach to increase mass spectrometry (MS) proteomics throughput is to shorten liquid chromatography (LC) separation times – for example, using rapid gradients, which are easily implemented with high-flow methods. Fast gradients (as short as 2–5 minutes) have been demonstrated in combination with DIA(100, 101). Specialized LC systems (e.g. Evosep) use pre-formed gradients to run over 100 injections per day with high reproducibility. However, decreasing chromatographic time reduces peptide separation, leading to less efficient separation. This may compromise proteome depth, especially for very short gradients, and may increase ionization suppression and interference. In practice, modern high-throughput workflows often seek a balance, using moderate gradient lengths or added separation dimensions (e.g. ion mobility) to preserve proteome coverage even as sample throughput increases. As the speed of new instruments increases, so does the coverage for short separations. Thus, newer instruments with fast scanning speeds enable throughputs of 100 single cells per day using label free DIA methods with relatively small tradeoffs in the depth of protein coverage. Yet, label free methods remain limited in achieving affordable single-cell proteomic analyses(102).

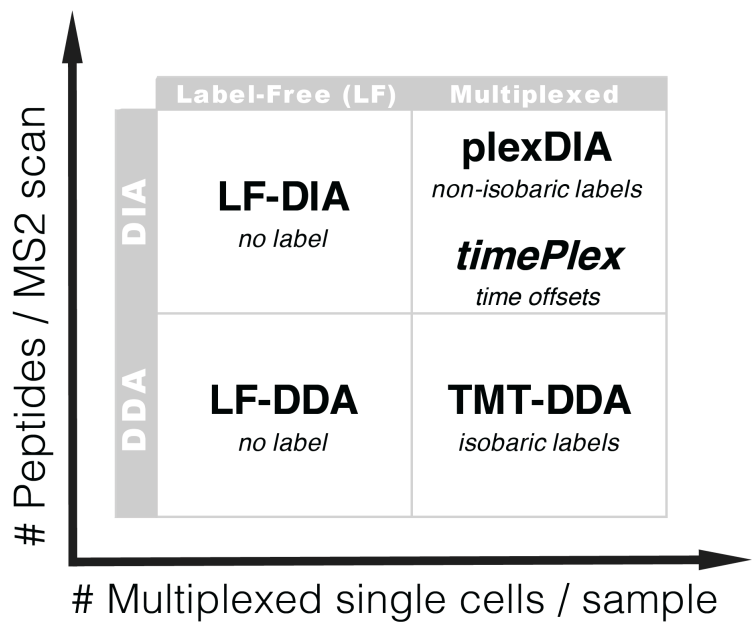

**Figure 2.** Categories of data acquisition methods based on the degree of parallelization. Data acquisition can be paralleled both across samples (single cells) and across proteins (number of simultaneously isolated and fragmented peptides). The figure is modified from ref.(103).

## Multiplexing samples with isobaric mass tags

Another major strategy to scale up proteomic throughput is sample multiplexing via isobaric labeling. Tandem mass tag (TMT) reagents (including TMTpro) allow pooling of many samples (up top 35 with the latest TMTpro reagents) and analyzing them in a single LC-MS/MS run. In single-cell proteomics, this approach (exemplified by SCoPE-MS/SCoPE2/pSCoPE) tags peptides from individual cells with different TMT labels and combines them, so that one MS run quantifies multiple single cells simultaneously. This parallelizes data acquisition at the sample level. Beyond increasing the throughput, isobaric mass tags allow the use of carriers (that pool and increase the detectable peptide fragments), which supports sequence identification. This benefit enabled some of the early studies that established the feasibility of single-cell MS proteomics(25, 28), and it continues to support analysis with older MS instruments.

In practice, TMT multiplexing has enabled the quantification of over 2,000 protein groups per cell. When combined with shorter gradients to increase throughout, it supports analyzing over 1,000 single cells per day while quantifying about 1,000 proteins per cell(15). The sensitivity and depth of coverage of these workflows are increased by using real-time instrument control (real-time search, prioritization) to help maximize MS/MS utilization for each run. A notable limitation, however, is that isobaric tags multiplex only at the sample level – the mass spectrometer still selects and fragments precursors sequentially, one at a time. As a consequence, conventional TMT-MS workflows remain limited by the MS/MS speed and cannot fully exploit modern MS instruments' capacity for parallel fragmentation. This can particularly limit the number of analyzed peptides from single cells since sensitive protein analysis benefits from longer ion accumulation times(16). Another limitation is the impact of co-isolation on quantification accuracy, which has been extensively discussed in the literature, e.g., as reviewed in ref.(104). The combination of isobaric labeling with SILAC can further increase the sample throughput though at the expense of limiting the number of MS/MS (and thus quantifiable peptides) that can be performed per single cell per unit time(105).

## Multiplexing both samples and peptides

Beyond multiplexing only samples (as enabled by isobaric mass tags), throughput can be further increased by multiplexing both samples and peptides. Such multiplexing can be achieved in the mass domain (plexDIA), in the time domain (timePlex), or in both domains simultaneously (Figure 2). These recent approaches chart a trajectory towards analyzing over 1,000 single cells per day without significant compromise in the accuracy and depth of proteomic analysis, as discussed below.

### Multiplexing in the mass domain

One approach for multiplexing both samples and peptides is multiplexed DIA (plexDIA), which uses non isobaric mass tags that encode each sample with a distinct mass/charge shift instead of identical isobaric masses. Originally demonstrated with mTRAQ mass tags(71), plexDIA was subsequently implemented with dimethyl tags(88), diethyl tags(106) and PSMtags(107). While plexDIA has been used in multiple biological single-cell proteomics studies(5, 89, 108), the utilized mass tags (mTRAQ) generally resulted in slightly lower coverage than the corresponding label free DIA can achieve. Yet, plexDIA tags may be optimized to improve sensitivity and amino acid sequence identification(86). To explore this potential, recent work screened a library of potential mass tags and introduced a tag termed PSMtags(107). By increasing the number of detected peptide fragments, PSMtags enhance sequence identification and support sequence identification for plexDIA proteomics. They also increase the throughput by supporting 5-plex (when using 4Da offsets) or 9-plex (when using 2Da offsets). Analyzing 9-plex samples of single-cell level standards on a 30 min gradient quantified over 3,000 proteins per sample. Extrapolating these results to 10min total analysis time per 9-plex single-cell sets suggests the potential of analyzing over 1,000 single cells per day per instrument without the tradeoffs of isobaric tags, reduced proteome coverage and quantification affected by coisolation.

### Multiplexing in the time domain

Another complementary approach for multiplexing both samples and peptides relies on multiplexing in the time domain and is accordingly named timePlex(69). In timePlex, samples are encoded by time offsets. A demonstrated implementation uses multiple staggered sample injections that introduce in the same MS instrument multiple samples with slight time offsets. For example, three single-cell samples can be injected sequentially a few minutes apart such that their active gradients interleave but remain distinguishable by the fixed time delays. Since multiplexing in the time and mass domains is multiplicative, combining 3-timePlex with 9-plex DIA results is 27-plex DIA, i.e., 27 samples per LC-MS run (3 time-offset injections × 9 mass-tagged samples). Crucially, this highly parallelized data acquisition increased the sample throughput without reducing the protein coverage per sample(69). This enabled over 500 samples per day to be processed on one instrument, with a clear path to exceed 1,000 cells/day by further increasing plex levels in the time and mass domains. These promising prospects for increasing the throughput require robust implementations before they can be broadly adopted.

### Analyzing highly multiplexed mass spectra

As discussed above, the approaches increasing throughput, include short separation times and parallelization in the time and mass domains. All of these approaches also increase the density of mass spectra, and thus the potential for overlap and interference between mass peaks of different ions. To address these challenges, software tools like DIA-NN and Spectronaut have developed dedicated modules to support plexDIA(93). These challenges motivated new tools, such as JMod, specifically developed to support multiplexing in both the mass and time domains(106). JMod performs joint modeling of overlapping signals to maintain quantification accuracy even when the isotopic envelopes of peptides from different multiplexed samples are overlapping. As the success of these tools attests, computational algorithms are an essential component supporting increasing proteomics throughput.

# Mechanistic Inference

A central promise of single-cell proteomics is the ability to help directly infer molecular mechanisms operating within cells by measuring the agents mediating these mechanisms – proteins and their interacting partners. Most biochemical interactions and signaling events involve proteins: enzymes modify substrates, receptors and kinases transmit signals, and multi-protein complexes execute functions. Therefore, if data for proteins is missing when modeling biological networks, they become unobserved confounders, hindering causal inference. Inferring such missing data from indirect proxies is unlikely to help much, as conditioning on confounders depends on high accuracy and precision measurements.

Causal inference from observational data in the presence of unobserved confers is fundamentally limited regardless of the algorithms used or the scale of the dataset(109). This problem is even harder if relying on RNA associations as statistical RNA associations often do not reflect direct biophysical interactions. Inferring direct causality from observational data alone is a notoriously hard problem: confounding variables and indirect effects can produce spurious associations. Even with infinite observational data, purely statistical approaches (akin to factor analysis) may fail to pinpoint true causal drivers due to fundamental ambiguities(109). In other words, correlation-based network inference can be fundamentally underdetermined – multiple causal models can explain the same correlation pattern, especially in high-dimensional biological data.

## Statistical association vs. biophysical causality

It is important to distinguish between two very different approaches to understanding causality in cell biology: statistical inference from observational data versus mechanistic modeling based on biophysical principles. Single-cell omics data (whether proteomic or transcriptomic) often invite the use of statistical or machine-learning methods to find associations – for example, inferring regulatory networks by correlating the abundance of RNA or proteins across cells(110–112). While such data-driven inference can suggest hypotheses, association is not causation. As a simple example, if protein X and protein Y are both elevated in a subset of cells, traditional correlation analysis might link them, but this could be because X and Y are co-regulated by some unseen factor Z rather than X causing Y or vice versa. Furthermore, the association between X and Y may be indirect, mediated by intermediate molecules that are highly context and condition dependent. Thus, such indirect associations are likely to vary across conditions, and models built upon them are unlikely to generalize.

Mechanistic, biophysical models take a different approach. Rather than relying on statistical associations alone, they incorporate known principles of chemistry and physics – such as binding affinities, enzymatic kinetics, and signaling pathways – to model how a system behaves. Crucially, the MS based technologies reviewed in the preceding sections have the potential to quantify such protein functions at scale, albeit this potential has not yet been realized. Thus, realizing this potential by comprehensive quantification of protein abundance and activities in single cells can better support the development of mechanistic models.

Biophysical models explicitly represent direct molecular interactions: for example, a model might include the reaction *Protein A phosphorylates Protein B*, with the associated rate constant. By building on such interactions, mechanistic models can simulate the dynamic behavior of a cell under various conditions. A key advantage is generalizability: a model grounded in true causal mechanisms is likely to hold even when conditions change, whereas a statistical model trained on one condition may break when extrapolated. For instance, a co-expression network might not predict what happens in a drug-treated cell, but a mechanistic model of the signaling pathway can, because it encodes causal relationships that persist under the intervention. Mechanistic models thus align with the idea of functional causality – they aim to explain *why* a change occurs by referencing an underlying physical interaction, not just *that* two variables are correlated.

## Conditioning on confounders

Developing mechanistic models requires rich data and direct measurements of the relevant variables, which is where single-cell proteomics can contribute. By providing measurements of actual effectors and signaling molecules, single-cell proteomic data make it easier to apply causal reasoning tools that go beyond associations. For example, if a computational method is trying to infer a protein regulatory network, having protein data for transcription factors (the direct regulators) will constrain the model to biologically plausible interactions (a modified transcription factor protein directly influencing target gene expression) rather than indirect transcript-transcript correlations. Early single-cell proteomic data have already provided evidence that the abundance of transcriptional factors (e.g., P53), unlike its transcript, is associated with the abundance of its target transcripts(26). Thus incorporation of protein data may help with the challenges of association-based inferences of transcriptional networks based in RNA-seq data(110).

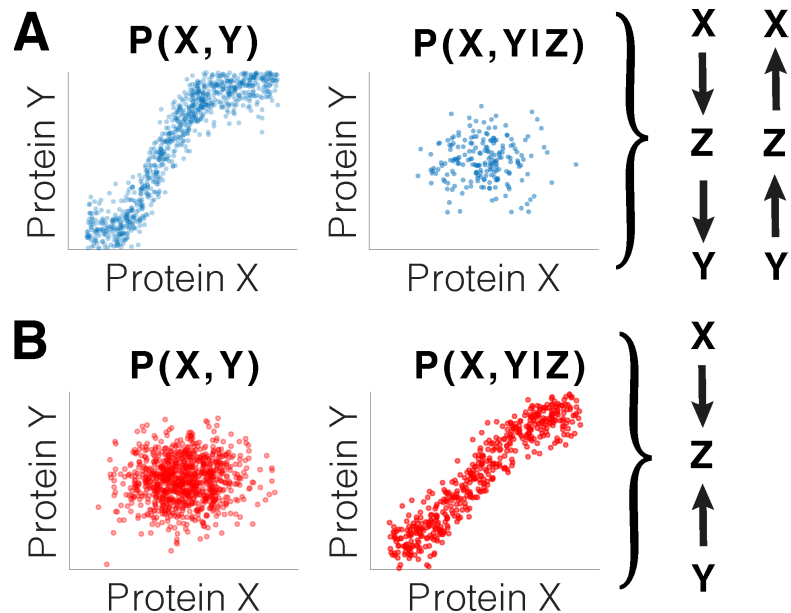

Figure 3. Distinguishing between direct and indirect regulation (A) Simulated data in which the joint variation of proteins X and Y. It illustrates that the interaction between X and Y is indirect, mediated by Z. (B) Another simulated example suggesting a collider model, where X and Y both influence Z, creating a conditional dependency between X and Y when accounting for Z. Arrows represent causal regulation (either activation or repression), and in the collider scenario, one arrow denotes positive regulation while the other indicates negative regulation. The figure is from ref.(111).

Transcending association-based inferences requires accounting for confounders, as illustrated in Figure 3. In causal inference terms, measuring intermediate nodes (proteins) in the cellular network can satisfy the conditions for identifying causal effects (e.g., by blocking hidden confounding paths). Indeed, one strategy to infer causation from observational data is to find negative controls or mediators that help isolate direct effects. When conditioning on confounded, it is essential that they are measured accurately since measurement noise can induce (rather than remove) spurious associations. Such accurate measurements are more feasible to achieve for proteins since MS methods can sample hundreds of copies per cell as opposed to the few copies sampled by single-cell RNA sequencing methods(26, 32). This type of causal inference relies solely on statistical dependencies observed across many single cells, but it critically depends on accurate single-cell measurements.

## Observing immediate responses to specific perturbations

Still, even with comprehensive single-cell protein data, purely observational inference has clear limitations. Definitive causal claims require observing the system response to targeted perturbations that modulate specific regulators. The perturbations need to be specific, affecting known proteins, rather than generic, e.g., treatment with a drug with unknown targets. The responses should be measured quickly after the perturbation since delayed responses are likely to include, even be dominated, by indirect secondary effects. One promising approach in this direction can be targeted protein degradation(113) followed by single-cell MS analysis soon after the targeted proteins are degraded.

In summary, statistical association-based inference can generate correlations and network hypotheses from single-cell data, but these are often confounded or incomplete. They generalize poorly outside of the training distributions. Biophysical models based principles provide a way to interpret data in terms of direct molecular interactions. They are more likely to reflect causation and generalize beyond the training distributions. Single-cell proteomics helps bridge the two, by supplying the rich, mechanistically relevant data needed to inform and constrain biophysical models. As the field progresses, we expect a convergence of data-driven inference with mechanistic knowledge, yielding predictive models of cells that are both data-grounded and principle-based.

# Outlook

Over the last decade, mass spectrometry analysis of proteins from single cells has transitioned from an area of skepticism to a robustly demonstrated and actively developing research field. The field has numerous clear technical objectives for progress, such as increasing the throughput and depth of protein analysis, and increasing the scope and accuracy of quantification of post translational modifications (PTMs). While variable search for PTMs could identify PTMs from the very beginning(28, 114), dedicated methods that can focus on the PTM by prioritized data acquisition(40) or spike-in peptide(87) can increase the scope and accuracy of PTM analysis. Still, more technological and methodological developments are needed for expanding the breadth and consistency of PTM quantification and protein activities towards enabling the mechanistic biophysical models discussed above. Advances towards these technical objectives can be realized by improvements of the fundamental analytical steps reviewed in this article, from mass tags and experimental designs to creative new data acquisition and analysis methods. These steps are advancing quickly, and their gains will continue to synergistically contribute towards increasing scope and accuracy of single-cell protein analysis, including of protein functions.

The biological applications of single-cell proteomics have proceeded from the very beginning to investigate cell fate transitions, macrophage polarization, cell lineage hierarchies, and drug resistance. Yet, the realization of the full potential for biologically driven research has generally lagged behind the technical capabilities of the technology and methods. Thus, a major opportunity for the field is to make the technology more accessible and better integrated with broader communities to enable diverse research objectives, including the mechanistic biophysical models advocated here.

Indeed, when combined with carefully designed perturbation experiments, single-cell proteomics may offer an exceptionally powerful route to mapping causal mechanisms: one can perturb a protein's activity and directly observe downstream protein-level effects in each cell, linking cause and effect with molecular resolution. This synergy of mechanistic measurement and perturbation-based validation is likely to be the gold standard for building reliable causal models in biology.

**Acknowledgements:** The work was funded by an NIGMS award R01GM144967 to N.S., and a MIRA award from the NIGMS of the NIH (R35GM148218) to N.S.

**Competing Interests:** N.S. is a founding director and CEO of Parallel Squared Technology Institute, which is a nonprofit research institute. N.S. Collaborates with Bruker and Thermo, which are manufacturers of MS instruments.